\documentclass[useAMS,usenatbib,usemath,usegraphicx,usedcolumn,amssymb]{mn2e}
\usepackage{amssymb}

\def\araa{{ARA\&A}}             
\def\apj{{ApJ}}                 
\def\apjl{{ApJ}}                
\def\aap{{A\&A}}                
\def\mnras{{MNRAS}}             
\def\nat{{Nature}}              

\title[A Compact Early-type Galaxy at $z = 0.6$ Under a Magnifying Lens]{A Compact Early-type Galaxy at $z = 0.6$ Under a Magnifying Lens: Evidence For Inside-out Growth}

\author[M. W. Auger, T. Treu, B. J. Brewer, and P. J. Marshall]{M. W. Auger$^{1}$\thanks{mauger@physics.ucsb.edu}, T. Treu$^{1}$, B. J. Brewer$^{1}$, P. J. Marshall$^{1,2}$\\
$^{1}$ Department of Physics, University of California, Santa Barbara, CA 93106, USA\\
$^{2}$ Kavli Institute for Particle Astrophysics and Cosmology, Stanford University, Stanford, CA 94305, USA} 

\begin{document}

\date{}

\pagerange{\pageref{firstpage}--\pageref{lastpage}} \pubyear{}

\maketitle

\begin{abstract}
We use Keck laser guide star adaptive optics imaging and exploit the magnifying effects of strong gravitational lensing (the effective resolution is FWHM $\approx 200$~pc) to investigate the sub-kpc scale of an intermediate-redshift ($z = 0.63$) massive early-type galaxy being lensed by a foreground early-type galaxy; we dub this class of strong gravitational lens systems EELs, e.g., early-type/early-type lenses. We find that the background source is massive (M$_* = 10^{10.9}$~M$_\odot$) and compact ($r_{\rm e} = 1.1$~kpc), and a two-component fit is required to model accurately the surface brightness distribution, including an extended low-surface-brightness component. This extended component may arise from the evolution of higher-redshift `red nuggets' or may already be in place at $z \sim 2$ but is unobservable due to cosmological surface brightness dimming.
\end{abstract}


\section{Introduction}
The hierarchical formation scenario of galaxy evolution suggests that galaxies grow via mergers with other galaxies. The early-type galaxies (ETGs) that are the products of these mergers follow a set of tight scaling relations, including a correlation between the sizes and stellar masses of ETGs at low redshifts \citep[e.g.,][]{shen,hyde}. However, it has recently been found that ETGs at high redshifts ($z \gtrsim 1.5$) show substantial deviations from this relation \citep[e.g.,][]{daddi,trujillo06,vanderwel,vanDokkum}; these `red nuggets' have effective radii between 2 and 5 times smaller for a given stellar mass than is expected from the local $r_e$-M$_*$ relation. Recent deep spectroscopic observations have found that the stellar velocity dispersions of these objects are consistent with the inferred sizes and stellar masses \citep[e.g.,][]{cappellari,vanDokkum09}, although there is some evidence that many galaxies have had their sizes systematically under-estimated \citep{cappellari}. Furthermore, \citet{newman} find that the sizes of ETGs at slightly lower redshifts than the red nuggets ($1 \lesssim z \lesssim 1.6$) are \emph{at most} a factor of two larger than ETGs in the local universe, which suggests that most of the size evolution of red nuggets -- if required -- occurs early in the lives of these galaxies.

Hierarchical growth through `dry' mergers should lead to most red nuggets growing substantially in size but not as quickly in mass, therefore moving these galaxies towards the local $r_e$-M$_*$ relation, although the tight scatter observed in the $r_e$-M$_*$ relation indicates that at most only half of the growth can be due to major mergers \citep{nipoti}. Additionally, up to 10\% of the red nuggets will not have experienced any major mergers to aid in this `puffing up' \citep{hopkins09a} and should thus exist at low redshifts. Even if some red nuggets exist locally, most will have disappeared, presumably evolving into the cores of massive early type galaxies \citep[e.g.,][]{hopkinsBundy,bezanson}. Several groups have attempted to find the surviving counterparts to these high-$z$ red nuggets at low redshifts with varying and sometimes discrepant results. \citet{valentinuzzi} examined the ETG population of the WINGS clusters and found that $\approx 20\%$ of these galaxies have properties similar to red nuggets and suggested that much of the observed evolution may be due to sample selection effects. Conversely, \citet{trujillo09} and \citet{taylor} searched the SDSS and found that the number density of red nuggets must decrease by a factor of $\sim 5000$ between $z \approx 2$ and $z \approx 0.1$. This discrepancy makes it unclear how many red nugget analogs exist in the local universe.

Most studies have investigated massive red nuggets in the distant ($z \gtrsim 1.5$) or local ($z \approx 0.1$) universe \citep[but see][]{stockton}. This is not due to a lack of interest in intermediate redshifts ($0.1 < z < 1$); indeed, the rate of evolution found by \citet{bernardi} suggests that the progenitors of $z = 0.1$ BCGs might still be red nuggets at $z < 1$, and \citet{hopkins10} suggest that tracing the evolution of the faint `outer envelope' of ETGs is a strong model discriminator. The problem with moving to intermediate redshifts has been in identifying compact red galaxies at these redshifts. For example, 0.5 kpc at $z = 0.5$ corresponds to 0.08$^{\prime\prime}$; such galaxies may be unresolved even in HST surveys.

We have implemented a novel approach to find and study intermediate-redshift analogs to the high-$z$ red nuggets: we use the SDSS spectroscopic database to find strong gravitational lens candidates that consist of an intermediate-redshift early-type galaxy being lensed by a lower-redshift early-type galaxy (we call these early-type/early-type lenses, or EELs). These natural telescopes provide a unique opportunity to identify and study red nuggets at $z \sim 0.6$, as high magnification is easier to achieve for compact sources \citep[e.g.,][]{marshall,newton} and ETGs are massive. Furthermore, the lensing magnification significantly increases the effective depth {\it and resolution} of observations \cite[e.g.,][]{treuReview}, opening the possibility to explore galaxies of unprecedented compactness.

This letter presents the discovery and first analysis of an EEL and demonstrates the unique utility of our approach. We present the data in Section 2, followed by our lensing and surface brightness fitting in Section 3, and conclude with a discussion in Section 4. All magnitudes are in the AB system, stellar masses are calculated assuming a Salpeter initial mass function \citep[e.g.,][]{augerIMF}, and we use a flat $\Lambda$CDM cosmology with $\Omega_{\rm m} = 0.3$ and $h = 0.7$.

\section{Observations and Data Reduction}
We have used the SDSS spectroscopic database and archive to identify spectra that appear to be composed of two early-type galaxy spectra at different redshifts and are therefore candidate strong gravitational lenses; this selection is similar to that employed by the Sloan Lens ACS \citep[SLACS;][]{slacsi} survey, although the SLACS survey looked for emission lines from the background objects and therefore selected late-type sources. The selection method essentially guarantees that the background galaxy is intrinsically luminous and therefore massive, as the spectroscopic continuum must be clearly visible in order to distinguish both the foreground and background redshifts.

We have observed 7 EEL candidates to date using NIRC2 with the laser guide star adaptive optics (LGS-AO) system on Keck II and have successfully confirmed that all of the observed candidates are strong gravitational lenses \citep{augerEELs}. Here we focus on the system SDSSJ1347-0101 with lens deflector redshift $z = 0.39$ and lens source redshift $z = 0.63$, which was observed on 21 April 2010 UT in clear skies using a tip-tilt star with magnitude $r = 16.8$ offset by 56\arcsec from the lens system. We used the NIRC2 wide camera (with pixel scale 0\farcs04 pixel$^{-1}$ and 0\farcs11 FWHM resolution during our observation) with the Kp filter and obtained 30 exposures of length 40s for a total on-source time of 1200s. Each exposure also includes a star located 17\arcsec from the lens which we use as the PSF star.

\begin{figure*}
 \begin{center}
 \includegraphics[width=0.98\textwidth,clip]{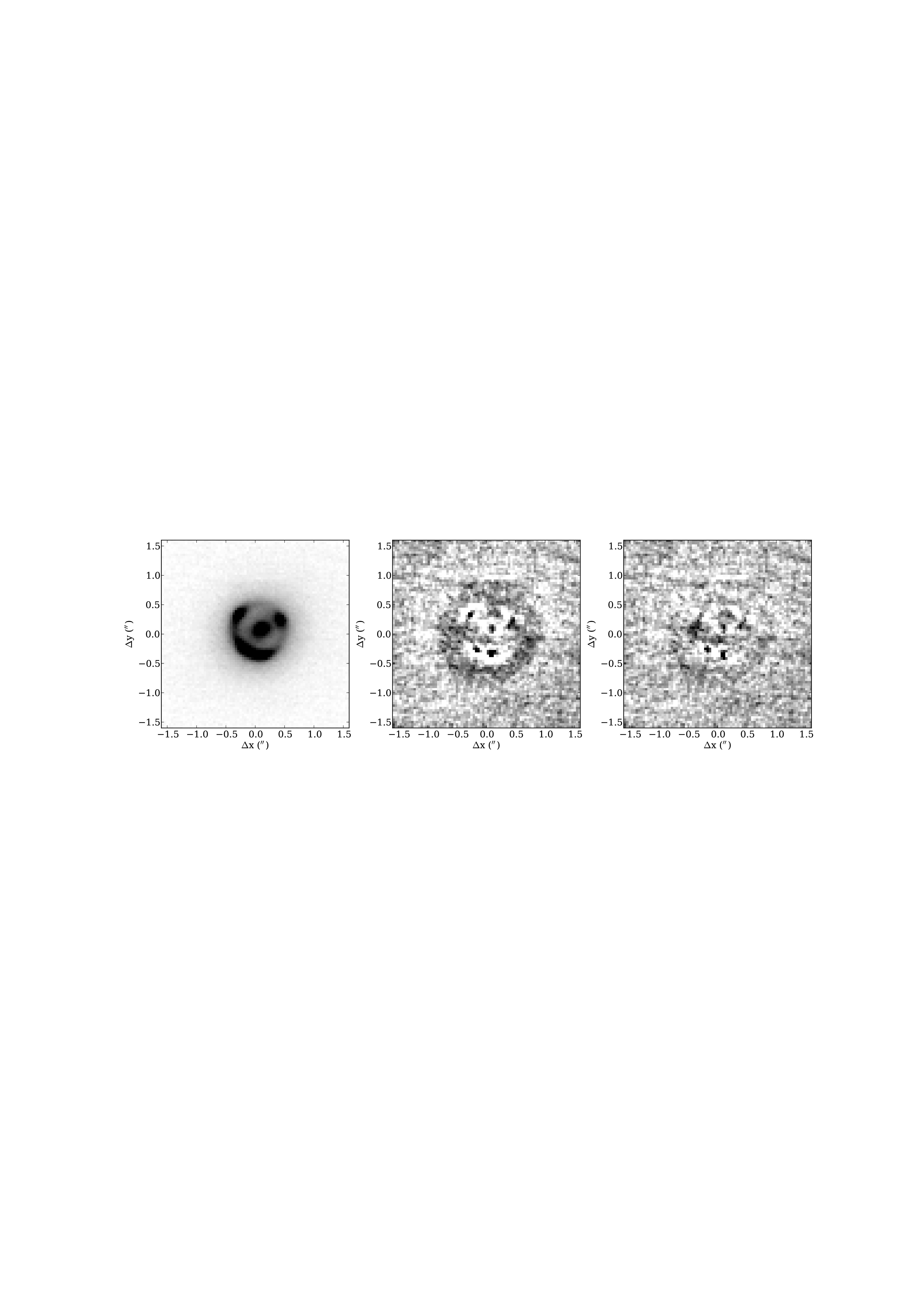}
 \caption{Keck LGS-AO imaging of the EEL SDSSJ1347-0101 (left). The image is well modelled using a singular isothermal ellipsoid mass distribution for the lensing mass, a single Sersic component for the lensing galaxy, and a two-component de Vaucouleurs and Sersic profile for the background source (right), while a single-component fit to the source leaves a ring of residual flux (centre); the scaling of the two residual plots is set to saturate at 3 times the background noise level. The Kp-band imaging reveals an observed source magnitude of 17.6 but this is corrected to 20.3 when the magnification from lensing is taken into account; the inferred stellar mass of the source is 10$^{10.9} {\rm M_\odot}$ and the half-light radius of the two-component surface brightness model is 0\farcs16 or 1.1 kpc at the redshift of the source.}
 \label{F_lens}
 \end{center}
\end{figure*}

The data were reduced by flatfielding with a sky flat created from the median of the science exposures with all sources masked, and a median background was then subtracted from each image. These images were resampled to a distortion-corrected frame and the 30 images were then registered using cross-correlation and pixel-based techniques which find sub-pixel offsets between the images. These offsets and the distortion model were then used to {\tt drizzle} the images onto a common output frame. The registration worked very well for the PSF star but we found that the distortion model left residual offsets of up to 0.7 pixels at the lens location, and we therefore re-registered the images with the PSF star masked to obtain an un-smeared final image of the lens system (Figure \ref{F_lens}).

\section{Mass and Surface Brightness Modelling}
We model the lens surface mass distribution as a singular isothermal ellipsoid \citep[SIE; e.g.,][]{kormann} and allow for external shear. The light of the lensing galaxy is modelled as an elliptical Sersic profile with Sersic index $n$ free to vary between 0.5 and 6, but we do not assume that the mass and the light of the lens are aligned. The background source was also initially modelled as a Sersic profile with Sersic index between 0.5 and 6. We convolve the models with an empirical PSF that was observed simultaneously with the lensing system and compare the convolved models with the data.

The best-fit model is obtained by using a Levenburg-Marquardt optimisation that minimises the $\chi^2$ of the difference between the data and the model. The data are generally well-described by the model, but we find that using a single Sersic component for the source results in a ring of residual flux left in the image (Figure \ref{F_lens}); the amplitude of these residuals is approximately 1.5 to 3 times the noise level. We also find that the data do not provide precise constraint on $n$ for the source (they only constrain $n$ to be large, i.e., $n > 3$) so we fix the source to have an elliptical de Vaucouleurs profile ($n = 4$). The residual flux is clearly not associated with the foreground deflector, as adding a second Sersic component does not improve the fit. However, adding a second component to the source surface brightness profile improves the fit dramatically, as shown in Figure \ref{F_lens}, and the per-pixel significance of the residuals around the ring is below the noise. The best-fit parameters for the lensing mass distribution and the light profiles for the source for both the one- and two-component models are given in Table \ref{T_parameters}.

\begin{table*}
 \begin{center}
  \caption{Best-fit parameters for the lens mass model and source light distribution of SDSSJ1347-0101. Columns: (1) Description of the source surface-brightness model; (2) lensing Einstein radius in arcsec; (3) lens model velocity dispersion in km~s$^{-1}$; (4) lens mass axis ratio $b/a$; (5) lens mass position angle in degrees; (6) external shear; (7) external shear position angle in degrees; (8) source magnitude; (9) source effective radius in arcsec; (10) source Sersic index. The errors on $m_{\rm Kp}$ are 0.15 mags and the errors on $r_{\rm e}$ are 10\%.}
  \begin{tabular}{@{}ld{2}cd{1}cd{2}cd{1}d{2}c@{}}
  \hline
  \multicolumn{1}{l}{Source Model} & \multicolumn{1}{r}{$r_{\rm Ein}$} & \multicolumn{1}{c}{$\sigma_{\rm SIE}$} & \multicolumn{1}{r}{$q_{\rm lens}$} & \multicolumn{1}{r}{PA$_{\rm lens}$} & \multicolumn{1}{r}{$\gamma_{\rm ext}$} & \multicolumn{1}{r}{PA$_{\gamma}$} & \multicolumn{1}{r}{$m_{\rm Kp}$} & \multicolumn{1}{r}{$r_{\rm e}$} & \multicolumn{1}{r}{$n$} \\
  \hline
  One-component  &  0.42  &  210  &  0.73  &  4  &  0.02  &  -62  &  20.6  &  0.07  &  4  \\
  Two-component  &  0.42  &  210  &  0.80  &  4  &  0.01  &  -39  &  20.3  &  0.16  & -- \\
  \ \ \ de Vaucouleurs & & & & & & &  21.3  &  0.04  &  4      \\
  \ \ \ Sersic        & & & & & & &  20.9  &  0.25  &  0.6   \\
\hline
\label{T_parameters}
\end{tabular}
\end{center}
\end{table*}

\begin{figure}
 \begin{center}
 \includegraphics[width=0.48\textwidth,clip]{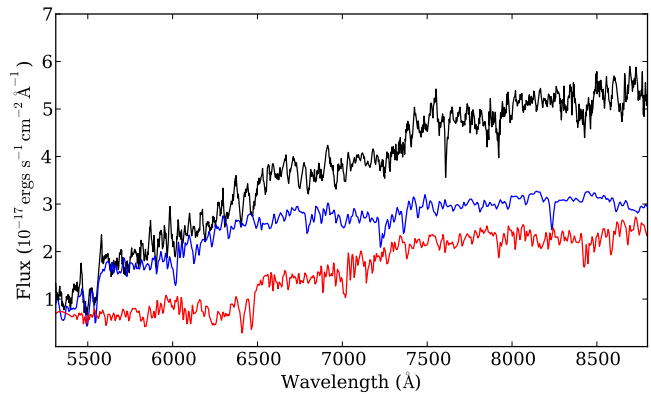}
 \caption{Smoothed SDSS spectrum of SDSSJ1347-0101 (black) and model template spectra for the lens (blue) and source (red). Ca K \& H lines from the background source are clearly visible in the smoothed spectrum, and we use the lens/source decomposition to determine a stellar velocity dispersion $\sigma_{\rm SDSS} = 203\pm42~{\rm km~s^{-1}}$ for the lens. Furthermore, we find that the stellar masses required to normalise the template spectra to match the observed spectrum are consistent with the masses derived from the Keck AO imaging (see Section 3).}
 \label{F_spectrum}
 \end{center}
\end{figure}

We find that the foreground galaxy is well-described by a single Sersic component for the light a SIE for the mass distribution. The SIE Einstein radius is 0\farcs42, which implies a lens velocity dispersion of 210$\pm15$~km~s$^{-1}$. This is consistent with the stellar velocity dispersion of 203$\pm42$~km~s$^{-1}$ that we obtain by fitting velocity-dispersion-broadened single-burst \citet{bc03} templates to the SDSS spectrum; Figure \ref{F_spectrum} shows the result of this fit, which we use to decompose the spectrum into lens and source components. The observed lens Kp magnitude is 17.7 and we use a \citet{bc03} instantaneous-burst galaxy template with formation redshift $z = 3$ and solar metallicity to determine rest-frame quantities. We find a K-band rest-frame magnitude of 18.4 and a luminosity of L$_{\rm K,lens} = 10^{11.4}~L_{\rm K,\odot}$, with a corresponding stellar mass M$_{\rm *,lens} = 10^{11.6}~{\rm M}_\odot$. We adopt error estimates of 0.1 dex on the luminosity and 0.2 dex on the stellar mass for both the lens and source.

The source has a total Kp magnitude of 20.3 (20.6 if a single-component is fit) and, again assuming a formation redshift $z = 3$ and solar metallicity, this implies that the rest-frame K-band magnitude is 21.0, the luminosity is L$_{\rm K,src} = 10^{10.8}~L_{\rm K,\odot}$, and the stellar mass is M$_{\rm *,src} = 10^{10.9}~{\rm M}_\odot$. The \emph{lensed} (i.e., observed) source magnitude is 17.6, and we therefore find a factor of 12 magnification due to lensing. The background source size is 0.5~kpc if a single de Vaucouleurs profile is assumed or 1.1~kpc if a two-component de Vaucouleurs and Sersic profile is used.

The SDSS spectrum allows us to verify that our stellar masses from the Keck AO-LGS photometry are robust. We fit \citet{bc03} templates to the foreground and background components of the spectrum to determine the appropriate normalisation of the templates and therefore the stellar mass of the flux observed by the SDSS spectroscopic fibre. The lens (foreground) stellar mass determined from the SDSS spectrum is $10^{11.1}~M_\odot$ which implies a 0.5 dex loss of flux due to the fibre aperture; this is consistent with what we find if we convolve our model with the SDSS point spread function (which has FWHM = 2\farcs2 for SDSSJ1347-0101) and integrate the flux within the SDSS fibre aperture. We perform the same procedure for the lensed (background) source and find that, when accounting for magnification and fibre losses, the source stellar mass inferred from the Keck LGS-AO photometry and the SDSS spectrum agree to within 0.2 dex, consistent with our assumed errors on the stellar mass.

\section{Discussion and Conclusions}
\begin{figure}
 \begin{center}
 \includegraphics[width=0.48\textwidth,clip]{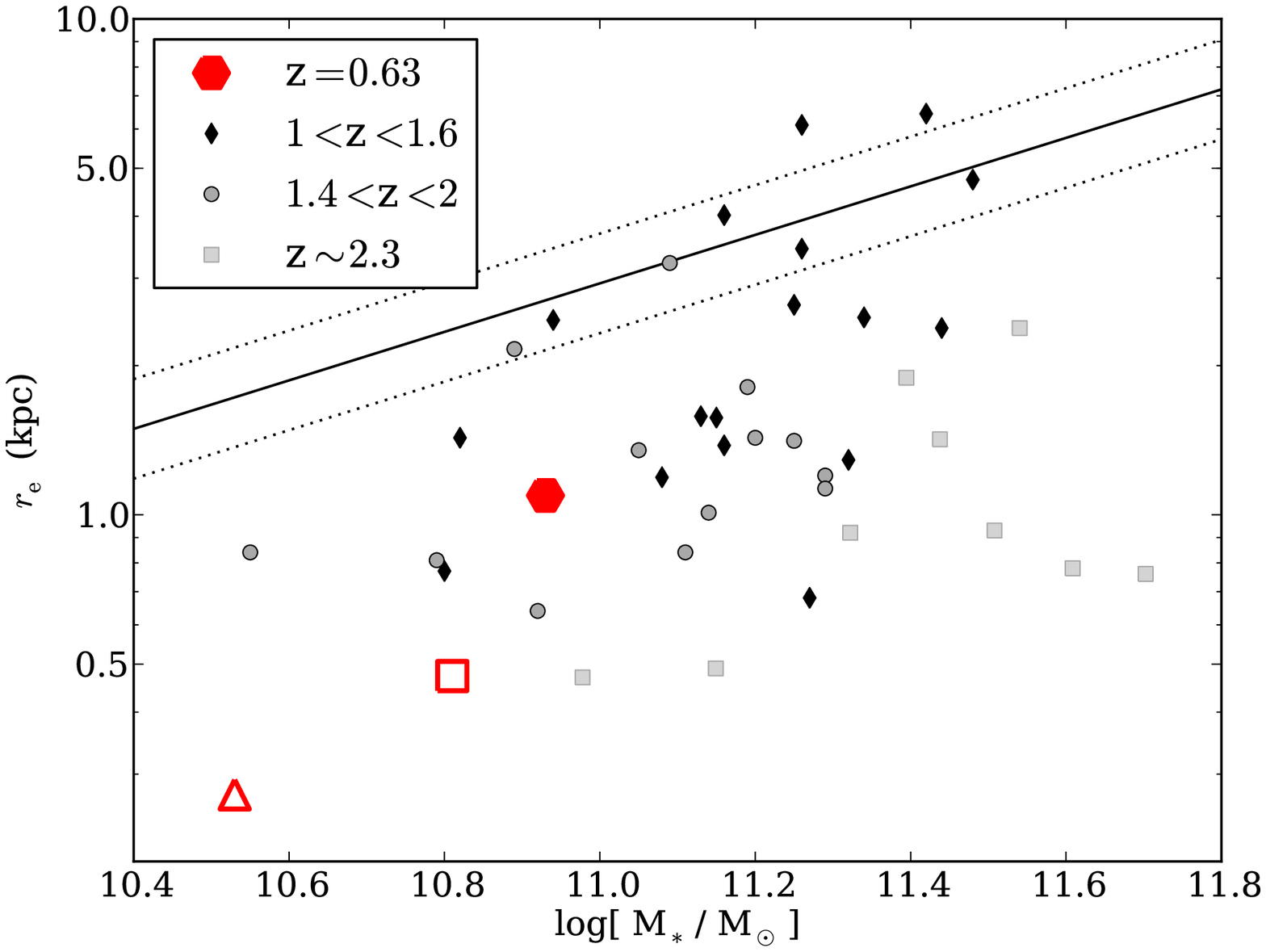}
 \caption{The size-mass relation for high-redshift ETGs; black diamonds are from \citet{newman}, grey circles are from \citet{cimatti}, and light-grey squares are from \citet{kriek} and \citet{vanDokkum}. SDSSJ1347-0101 (red points; the square is for a single-component de Vaucouleurs fit, the hexagon is for the two-component fit, and the triangle is the de Vaucouleurs component of the two-component fit) has properties consistent with the high-redshift objects but is a significant outlier from the local-universe size-mass relation \citep[black line with 1-$\sigma$ scatter indicated by dotted lines;][]{hyde}.}
 \label{F_sizeMass}
 \end{center}
\end{figure}

The background source of SDSSJ1347-0101 is a clear outlier from the size-mass relationship of local ETGs and has properties that are similar to high-redshift ETGs (Figure \ref{F_sizeMass}). However, the inferred size varies by a factor of two depending on whether a one- or two-component surface brightness model is fit to the galaxy; our imaging data clearly indicate that the two-component model is required to describe the light distribution of SDSSJ1347-0101. Larger samples of intermediate-redshift objects are need to make any strong claims, but we note that the extended `wing' of flux from the background source of SDSSJ1347-0101 may help alleviate the tension between the very compact high-redshift galaxies and the apparent lack of local-universe counterparts \citep[e.g.,][]{trujillo09,taylor} in two ways.

First, single-component fits may be poor models for the true surface brightness distribution of compact ETGs \citep[e.g.,][]{stockton}, and observations of high-redshift galaxies may lack the sensitivity required to characterise low-surface-brightness extended components of red nuggets. This is unlikely to fully resolve the issue, however, since a factor of two size difference is not enough to move all of the high-redshift objects to the local size-mass relation. We also note that deep \emph{HST} imaging of the high-redshift galaxies does not indicate the presence of a significant second component for most red nuggets \citep[e.g.,][]{damjanov}.

\begin{figure}
 \begin{center}
 \includegraphics[width=0.48\textwidth,clip]{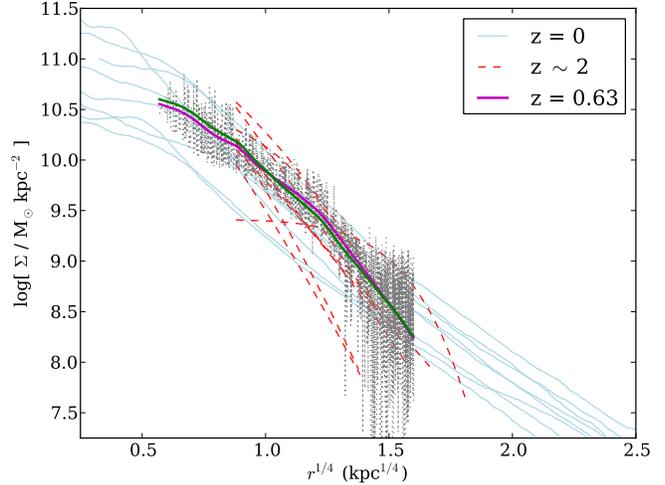}
 \caption{The observed surface density profile of the background source of SDSSJ1347-0101 (dotted grey line) and best-fit one-component (green) and two-component (purple) models plotted from the HWHM of the effective resolution to the surface brightness limit of our observations. We also plot the surface brightness profiles for massive Virgo ellipticals from \citet{hopkinsBundy} rescaled to a Salpeter IMF (blue lines) and for the $z \sim 2$ red nuggets (red lines) from \citet{vanDokkum}. The single-component model is unable to fit the low-surface-density tail and overfits the central region to compensate, as is seen in Figure 1.}
 \label{F_sb_profile}
 \end{center}
\end{figure}

Alternatively, the low-surface-brightness components of intermediate-redshift red nuggets might be due to evolution of the higher-redshift red nuggets. \citet{hopkinsBundy} show that massive ETGs in the local universe have extended low surface brightness components but the central surface mass density of these galaxies is similar to the central surface mass density of the red nuggets, indicating that the red nuggets may `grow' extended wings of low-surface-brightness material. We find that SDSSJ1347-0101 is consistent with this scenario and has a peak (PSF-deconvolved) surface mass density of $10^{11.2} {\rm M_\odot kpc^{-2}}$, similar to both low- and high-redshift massive ETGs (Figure \ref{F_sb_profile}).

\citet{stockton} used 9-band photometry from SDSS and UKIDSS to identify compact objects that have colours consistent with ETGs at redshifts $0.4 \lesssim z \lesssim 0.6$. They found 14 compact massive galaxy candidates in an area of $\sim$200 square degrees and followed up 11 the them; 7 of these 11 were spectroscopically found to be galaxies and two were confirmed to be compact ($r_e \approx 1$ kpc) using LGS-AO imaging with NIRC2. These intermediate-redshift, massive compact galaxies also require low-surface-brightness components to adequately describe their surface brightness distributions, and this may indicate that lower-redshift compact massive galaxies require two-component models in general.

A larger number of intermediate-redshift red nuggets must be investigated to determine quantitatively their density, test the degree to which different physical mechanisms affect the evolution of the high-redshift red nuggets, and characterise the scatter in the size-mass relationship of these objects \citep[e.g.,][]{nipoti}. Depending on the slope of the size-luminosity function, lensed sources may be more compact than non-lensed counterparts due to the increased cross-section for lensing of high-surface-brightness sources \citep[e.g.,][]{newton}. Our Keck adaptive-optics imaging has been extremely successful at confirming the lensing nature of EEL candidates discovered in the SDSS spectroscopic database. However, there are several potential drawbacks to this approach, all of which are surmountable.

The first is the non-trivial selection function which makes calculating precise number densities, for example, difficult. \citet{dobler} have characterised the selection function for the SLACS survey, and a similar approach will yield the appropriate volume in which we search for EELs. We can nevertheless obtain an order of magnitude estimate of the number density of EEL sources by making some approximations. We find that roughly one out of every 1000 SDSS spectra (which have circular apertures with radius 1\farcs5) contain a high-quality EEL candidate. If we assume that we are uniformly sensitive to sources between $0.3 < z < 0.7$ (this is an approximation; we neglect magnification bias and the actual selection is biased towards the higher redshift end of this interval due to the lensing cross section), this yields an EEL source density of $1.3\times10^{-3}$~Mpc$^{-3}$. This is a factor of a few smaller than the density of massive ETGs at these redshifts \citep[e.g.,][]{bundy}, indicating that EELs sources are not an extreme subset of the overall population For comparison, the number density of EELs sources is slightly higher than high-redshift red nuggets \citep[e.g.,][]{taylor} and is several orders of magnitude larger than that of intermediate-redshift red nuggets inferred from the \citet{stockton} search (they find one object per 100 square degrees at $0.4 < z < 0.6$ while we find one EEL in every 2 square arcminutes at $0.3 < z < 0.7$). This difference is due in part to our higher sensitivity (and therefore lower stellar mass limit), but it also indicates that the selection function of \citet{stockton} is much more restrictive and suggests that the EEL source population may be more representative of the overall population of compact red galaxies at intermediate redshifts.

Two other drawbacks of our approach are related to the use of AO imaging to confirm and analyse the sources. The requirement of having a bright tip-tilt star within one arcminute of the lens substantially limits the ability to follow-up targets and therefore decreases the possible sample size. Additionally, accurately modelling the sources requires a good model for the AO PSF; in principle this can be obtained by observing stars off-axis from the science observation, although we have found that results are significantly more robust when the PSF star is observed simultaneously with the lens, as in the present case of SDSSJ1347-0101. We note, however, that space-based imaging of EEL candidates alleviates both of these concerns due to the stable PSF from space and the ability to observe any part of the sky.

Finally, we note that the strong lensing magnification increases the observed flux from the lensed source and therefore makes it easier to obtain spectroscopic follow-up of the background source (although the presence of the foreground lensing galaxy complicates the interpretation of the spectra). Our program therefore provides an efficient means of exploring the size-mass and mass-velocity dispersion relations of massive compact galaxies at intermediate redshifts, and we may be able to use spectroscopy to explore the stellar populations of the sources in detail to investigate evidence of a younger population of stars associated with the extended low-surface-brightness profile.

\vskip 0.1in
\noindent\textbf{ACKNOWLEDGEMENTS}

\noindent We thank K.~Bundy for useful discussions and the referee for a constructive report. TT acknowledges support from the NSF through CAREER award NSF-0642621 and from the Packard Foundation through a Packard Fellowship. PJM acknowledges the Kavli Foundation for support in the form of a research fellowship. This Letter includes data obtained at the W.~M.~Keck Observatory, operated by Caltech, the University of California, and NASA, and made possible by the W.~M.~Keck Foundation. We recognise the very significant cultural role and reverence that the summit of Mauna Kea has always had within the indigenous Hawaiian community. This work has made extensive use of the SDSS database.

\label{lastpage}


\begin{thebibliography}{}

\bibitem[Auger et al.(2010a)]{augerIMF} Auger, M.~W., Treu, T., Gavazzi, R., Bolton, A.~S., Koopmans, L.~V.~E., \& Marshall, P.~J.\ 2010a, \apjl, 721, L163

\bibitem[Auger et al.(2010b)]{augerEELs} Auger, M. W., et al.\ 2010b, in preparation

\bibitem[Bernardi(2009)]{bernardi} Bernardi, M.\ 2009, \mnras, 395, 1491

\bibitem[Bezanson et al.(2009)]{bezanson} Bezanson, R., van Dokkum, P.~G., Tal, T., Marchesini, D., Kriek, M., Franx, M., \& Coppi, P.\ 2009, \apj, 697, 1290

\bibitem[Bolton et al.(2006)]{slacsi} Bolton, A.~S., Burles, S., Koopmans, L.~V.~E., Treu, T., \& Moustakas, L.~A.\ 2006, \apj, 638, 703

\bibitem[Bruzual \& Charlot(2003)]{bc03} Bruzual, G., \& Charlot, S.\ 2003, \mnras, 344, 1000

\bibitem[Bundy et al.(2006)]{bundy} Bundy, K., et al.\ 2006, \apj, 651, 120

\bibitem[Cappellari et al.(2009)]{cappellari} Cappellari, M., et al.\ 2009, \apjl, 704, L34

\bibitem[Cimatti et al.(2008)]{cimatti} Cimatti, A., et al.\ 2008, \aap, 482, 21 
\bibitem[Damjanov et al.(2009)]{damjanov} Damjanov, I., et al.\ 2009, \apj, 695, 101 

\bibitem[Daddi et al.(2005)]{daddi} Daddi, E., et al.\ 2005, \apj, 626, 680

\bibitem[Dobler et al.(2008)]{dobler} Dobler, G., Keeton, C.~R., Bolton, A.~S., \& Burles, S.\ 2008, \apj, 685, 57 

\bibitem[Hopkins et al.(2009a)]{hopkins09a} Hopkins, P.~F., Hernquist, L., Cox, T.~J., Keres, D., \& Wuyts, S.\ 2009a, \apj, 691, 1424

\bibitem[Hopkins et al.(2009b)]{hopkinsBundy} Hopkins, P.~F., Bundy, K., Murray, N., Quataert, E., Lauer, T.~R., \& Ma, C.-P.\ 2009b, \mnras, 398, 898 

\bibitem[Hopkins et al.(2010)]{hopkins10} Hopkins, P.~F., Bundy, K., Hernquist, L., Wuyts, S., \& Cox, T.~J.\ 2010, \mnras, 401, 1099

\bibitem[Hyde \& Bernardi(2009)]{hyde} Hyde, J.~B., \& Bernardi, M.\ 2009, \mnras, 394, 1978

\bibitem[Kormann et al.(1994)]{kormann} Kormann, R., Schneider, P., \& Bartelmann, M.\ 1994, \aap, 284, 285

\bibitem[Kriek et al.(2008)]{kriek} Kriek, M., et al.\ 2008, \apj, 677, 219

\bibitem[Marshall et al.(2007)]{marshall} Marshall, P.~J., et al.\ 2007, \apj, 671, 1196

\bibitem[Newman et al.(2010)]{newman} Newman, A.~B., Ellis, R.~S., Treu, T., \& Bundy, K.\ 2010, \apjl, 717, L103 

\bibitem[Newton et al.(2010)]{newton} Newton, E.~R., et al.\ 2010, submitted to \apj

\bibitem[Nipoti et al.(2009)]{nipoti} Nipoti, C., Treu, T., Auger, M.~W., \& Bolton, A.~S.\ 2009, \apjl, 706, L86 

\bibitem[Shen et al.(2003)]{shen} Shen, S., Mo, H.~J., White, S.~D.~M., Blanton, M.~R., Kauffmann, G., Voges, W., Brinkmann, J., \& Csabai, I.\ 2003, \mnras, 343, 978

\bibitem[Stockton et al.(2010)]{stockton} Stockton, A., Shih, H.-Y., \& Larson, K.\ 2010, \apjl, 709, L5

\bibitem[Taylor et al.(2010)]{taylor} Taylor, E.~N, Franx, M., Glazebrook, K., Brinchmann, J., van der Wel, A., \& van Dokkum, P.~G 2010, \apj, 720, 723 

\bibitem[Treu(2010)]{treuReview} Treu, T.\ 2010, \araa, 48, 87 

\bibitem[Trujillo et al.(2006)]{trujillo06} Trujillo, I., et al.\ 2006, \mnras, 373, L36 

\bibitem[Trujillo et al.(2009)]{trujillo09} Trujillo, I., Cenarro, A.~J., de Lorenzo-C{\'a}ceres, A., Vazdekis, A., de la Rosa, I.~G., \& Cava, A.\ 2009, \apjl, 692, L118   

\bibitem[Valentinuzzi et al.(2010)]{valentinuzzi} Valentinuzzi, T., et al.\ 2010, \apj, 712, 226  

\bibitem[van der Wel et al.(2008)]{vanderwel} van der Wel, A., Holden, B.~P., Zirm, A.~W., Franx, M., Rettura, A., Illingworth, G.~D., \& Ford, H.~C.\ 2008, \apj, 688, 48 

\bibitem[van Dokkum et al.(2008)]{vanDokkum} van Dokkum, P.~G., et al.\ 2008, \apjl, 677, L5 

\bibitem[van Dokkum et al.(2009)]{vanDokkum09} van Dokkum, P.~G., Kriek, M., \& Franx, M.\ 2009, \nat, 460, 717 

\end{thebibliography}
\end{document}